\documentclass[12pt]{article}
\usepackage{amsfonts,amssymb,amsmath,latexsym}
\newcommand{\be}{\begin{equation}}
\newcommand{\ee}{\end{equation}}
\newcommand{\bea}{\begin{eqnarray}}
\newcommand{\eea}{\end{eqnarray}}

\newcommand{\ba}{\begin{array}}
\newcommand{\ea}{\end{array}}

\newcommand{\bt}{\begin{tabular}}
\newcommand{\et}{\end{tabular}}

\newcommand{\fr}{\frac}
\newcommand{\ci}{\cite}
\newcommand{\cl}{\centerline}
\newcommand{\bs}{\bigskip}
\newcommand{\vs}{\vspace}

\newcommand{\en}{\eqno}

\newcommand{\bbib}{\begin{thebibliography}}
\newcommand{\ebib}{\end{thebibliography}}

\newcommand{\mbb}{\mathbb}

\begin{document}

\titlepage
\hspace{5cm} {\it LANDAU INSTITUTE preprint 03/02/00}
\vspace{1.5cm}

\centerline{\large{\bf D-DIMENSIONAL CONFORMAL }}

\bs

\centerline{\large{\bf $\sigma$-MODELS AND TOPOLOGICAL EXCITATIONS}}


\vs{1cm}

\cl{\bf S.A.Bulgadaev}

\bs

\cl{\it L.D.Landau Institute, Kosyghin Str.2, Moscow, Russia,117334}
\vs{1cm}

\cl{Abstract}

\vs{1cm}

The D-dimensional conformal nonlinear sigma-models (NSM) are constructed.
It is shown that the NSM on spaces with $\pi_{D-1} = \mbb{Z}$ have
the topological solutions of a "hedgehog"
and "anti-hedgehog" type with logarithmic energies.
For spaces with $\pi_{D} \ne 0$ they have also the topological excitations
with finite energies of instanton types.

\newpage


\cl{\bf I. Introduction.}

\bs

It is known the topological excitations (TE) with logarithmically divergent
energy play important role
in low-dimensional systems with $D \le 2$, where they induce the
topological phase transitions (TPT) [1-4].
As can be shown for the existence of such topological excitations
one needs the configurations of the
boundary values of field, determining a homotopically
nontrivial mapping of the physical space boundary
$\partial \mbb {R}^D = S^{D-1}$ into the vacuum manifold ${\cal M}$
of the internal space ${\cal R}^N,$
and a scale invariant action \ci{5}.
An existence of such configurations depends on the nontriviality of
the corresponding homotopic group $\pi_{D-1}({\cal M}).$
Due to this property the TPT
admits another, topology changing, interpretation \ci{5}.
In the massive (high-temperature) phase the freely existing
TE conserve an information about the compactness and topology of the
vacuum manifold ${\cal M}$ and give the short range correlations.
In the massless (low-temperature) phase the TE are
bounded due to logarithmic
interaction into neutral dipoles.
This gives a lost of information about the compactness and topology of
${\cal M}$, at least, {\it  on large scales}, and the theory behaves as a
{\it free, noncompact, conformal} one. Note that under such TPT there is
no real topological reconstruction of ${\cal M}$ at the phase transition point.
The TPT gives only {\it statistical decompactification} of the theory.

After construction of the theory of low-dimensional TPTs a question
has naturally  arised
about a possibility of such TE and TPT in systems with dimensions $D>2.$
However, in higher dimensional systems the main efforts were devoted
to the discovery
of the TE with {\it finite energy} [6-9]. All such
excitations give finite contribution to the partition function,
but cannot induce a PT similar to the TPT, since the latter is induced by
TE with {\it logarithmically divergent energy}.

Unfortunately, the TE with logarithmic energies do not appear
in usual higher dimensional theories with action bilinear on gradient.
Instead it has been only shown that the partition function
of "particles"  with
logarithmic interaction can appear in some cases. For example,
such partition function
appears as a measure on the parameter (or moduli) space of instantons
in 4-dimensional Yang-Mills theory (the so called "merons") \ci{10},
and in 2-dimensional nonlinear $\sigma$-model (NSM)\ci{11}.
But these "particles" were not the TE.

Recently it was shown that the TE with logarithmic energy can exist
in 3D conformal (or the van der Waals) nonlinear $\sigma$-model (NSM) \ci{12}.
In this paper we consider a possibility of existence of TE
with logarithmical
energies in higher ($D>3$) dimensional systems, whose behaviour
at large scales can be described by generalized nonlinear $\sigma$-models.
It will be shown that an existence of such excitations
is intimately related  with a conformal symmetry of the models.
It appears also that these models admit another TE with finite energies
similar to the instantons.

\bs

\cl{\bf II. D-dimensional conformal invariant nonlinear $\sigma$-models}


\bs

Here we will consider only the models of nonlinear $\sigma$-model type.
More accurate analysis of these models
has shown that an existence of
topological excitations with discrete topological charges and logarithmic
energy puts over the following properties on NSM:

1) their homotopical group
$\pi_{D-1}({\cal M})$ must be nontrivial and abelian discrete,

2) a conformal invariance at classical level.

The first property
permits some ambiguity in a dimension and a form of ${\cal M}$.
while the second one
defines a form of the NSM action ${\cal S}$
almost uniquely in arbitrary dimensions.

A general expression for action ${\cal S}$ of $D$-dimensional generalized
nonlinear $\sigma$-models,
admitting nonlocal ones, can be represented in the next form
$$
{\cal S} = \fr{1}{2\alpha} \int d^D x d^D x' \psi^a(x)
\boxtimes_{ab}^{(D)}(\psi|x,x') \psi^b(x'),\quad a,b=1,...,n
\en(1)
$$
where $\psi \in {\cal M},$ the  manifold of degenerate vacuum states of the
model, and $n$ is its dimension. The form of the kernel $\boxtimes$ depends
on model. If the structures of the internal and physical spaces do not depend
on each other and the latter space is homogeneous, then $\boxtimes$ can
be decomposed
$$
\boxtimes^{(D)}_{ab}(\psi|x,x') = g_{ab}(\psi(x),\psi(x')) \Box_D (x-x'),
\en(2)
$$
where $g_{ab}$ is some two-point metric function on ${\cal M}.$
For local models an expression for $\boxtimes$ can be defined in
terms of manifold ${\cal M}$ only
$$
\boxtimes(\psi|x,x') = g_{ab}(\psi)\boxtimes \delta(x-x')
\eqno(3)
$$
If the manifold
${\cal M}$ can be embedded in Euclidean vector space ${\mbb{R}}^{N(n)}$
with dimension $N(n),$ depending on $n,$ then one can use instead of
$g_{ab}(\psi,\psi')$ an usual Euclidean metric
$$g_{ab} = \delta_{ab}, \quad a,b=1,...,N.
\en(4)
$$
and the  constraints defining ${\cal M}.$
Due to the first condition we must consider the manifolds with a discrete
abelian homotopic group $\pi_{D-1}({\cal M}).$
The spheres $S^{D-1}$ are the simplest among them with
$\pi_{D-1}({\cal M})= \mbb {Z}.$
Then  $N(D-1)= D$, and
$$\psi^a = n^a, \quad a = 1,...,D, \quad ({\bf n})^2 = 1.
\en(5)
$$
where ${\bf n}(x)$ is a field of unit vectors in internal space ${\cal R}^D.$
Since $N = D$  this internal space can be identified
with a physical space.
Below we confine ourselves by the simplest manifolds, the spheres.
A possible generalizations will be shortly discussed in conclusion.

For analyzing of the possible actions (and kernels)
it is more convenient to write ${\cal S}$ and $\Box_D$ in the momentum space
$$
{\cal S} = \fr{1}{2\alpha} \int d^D x d^D x' ({\bf n}(x){\bf n}(x'))
\Box_D(x-x') =
\fr{1}{2\alpha} \int \fr{d^D k}{(2\pi)^D} |n(k)|^2 \Box_D(k)
\en(6)
$$
For asymptotical scale invariance at large scales the kernel $\Box_D$ must
have  the next behaviour at small $k$
$$
\Box_D (k)\simeq |k|^D (1 + a_1 (ka) + ...) = |k|^D f(ka)
\eqno(7)
$$
where $a$ is a UV cut-off parameter, $f(ka)$ is some regularizing function
with the next asymptotics
$$
f(ka) = 1 + a_1 ka + ..., \quad ka \to 0,
\quad f(ka) \to 0, \quad ka \to \infty.
\en(8)
$$
The kernel $\Box_D$ generalizes an usual  local and
conformal kernel of two-dimensional $\sigma$-model
$$
\Box_2(k)\equiv \Box(k) = - (\partial)^2(k) = k^2
\eqno(9)
$$
For this reason $\Box_D$ can be also named as a $|\partial|^D$ kernel.
From (9) it follows that in even dimensional spaces $\mbb {R}^{2s}$
the kernel $\Box_{2s}$ is a local one
$$
\Box_{2s} = (-1)^s ((\partial)^2)^s.
\en(10)
$$
In odd dimensions  $\Box_D$ is always a nonlocal one with a large-scale
asymptotics
$$
\Box_D(x)|_{x \gg a} \simeq  A_D/|x|^{2D}, \quad
$$
$$
A_D = \fr{S_{D-2} 2^{2D-1} \Gamma(\fr{1}{2})\Gamma(\fr{D-1}{2})\Gamma(D)}
{(2\pi)^D \Gamma(-\fr{D}{2})} =
\fr{2^D \Gamma(D)}{\pi^{D/2}\Gamma(-\fr{D}{2})},
\quad D=2s+1,
\en(11)
$$
where $S_{D-2} = \fr{2 \pi^{\fr{D-1}{2}}}{\Gamma(\fr{D-1}{2})}$ is a
volume of the $(D-2)$-dimensional unit sphere.
Such kernels appear often in physics.
The most known and important
cases correspond to the Calogero-Sutherland-Dyson \ci{13} and
to the Caldeira-Leggett kernel
($D=1$) \ci{14} and to the van der Waals potential ($D=3$) \ci{12}.

There are analogous nonlocal kernels in even-dimensional spaces
$$
\Box_{2s}(x) \sim 1/x^{2s},
\en(12)
$$
but their Fourier-images contain an additional
logarithmic factor
$$
\Box_{2s}(k) \sim k^{2s}\ln (k/k_0)
\en(13)
$$
where $k_0$ is some UV cut-off parameter, regularizing a kernel (11)
at small scales.
This factor breaks some important properties of the model.
Further we will take in even-dimensional spaces only a local kernel.

As is well known  a  conformal group in the higher-dimensional ($D>2$)
spaces is finite-dimensional \ci{15}. Its main nontrivial transformation
is an inversion transformation
$$
x^i \to x^i/r^2, \quad r = |{\bf x}|.
\en(14)
$$
The conformal invariance of ${\cal S}$ with a kernel (11)
follows from the next transformation properties of the kernel $\Box_D$
under conformal transformation (14)
(the field ${\bf n}(x)$ and a coupling constant
$\alpha$ are dimensionless)
$$
x_i \to  x'_i = x_i/r^2, \quad r \to r' = 1/r, \quad x_i/r = x'_i/r',
$$
$$
d^D x \to d^D x/|{\bf x}|^{2D},\quad
\fr{1}{|{\bf x}_1-{\bf x}_2|^{2D}} \to
\fr{|{\bf x}_1|^{2D} \,|{\bf x}_2|^{2D}}{|{\bf x}_1-{\bf x}_2|^{2D}},
$$
and, consequently,
$$
{\cal S} = \fr{A_D}{2\alpha} \int d^D x_1 d^D x_2 \;
\fr{({\bf n}_1{\bf n}_2)}{|{\bf x}_1-{\bf x}_2|^{2D}}
\en(15)
$$
is invariant and dimensionless.
Thus the action (15) with a kernel (11) can be named a  D-dimensional
{\it conformal} NSM. Note that this invariance takes place for even dimensions
also. Strictly speaking, a conformal invariance takes
place only at large scales, since a kernel (11) needs some regularization
at small distancies, which can break this invariance.

The corresponding Euler - Lagrange equation has a form
$$
\int \Box_D(x-x') {\bf n}(x')d^D x' -
{\bf n}(x) \int ({\bf n}(x){\bf n}(x')) \Box_D(x-x') d^D x' = 0.
\en(16)
$$
The Green function $G^D(x)$ of the conformal kernel
$\Box_D(x)$ can be defined by next equation
$$
\int \Box_D(x-x'') G^D(x''-x') d^D x'' = \delta (x-x')
\en(17)
$$
It has the following form
$$
G^D(x) = \Box_D^{-1} = |\partial|^{-D} =
\left.\int \fr{d^D k}{(2\pi)^D}\fr{e^{i({\bf k}{\bf x})}}
{\Box_D(k)} \right.
\en(18)
$$
At large scales $G^D(x)$  has a logarithmic asymptotic behaviour
$$
G^D(x)|_{r \gg a} \simeq
- B_D \ln (r/R), \quad
$$
$$
B_D = \fr{S_{D-2}
\Gamma(\fr{1}{2})\Gamma(\fr{D-1}{2})}{(2\pi)^D \Gamma(\fr{D}{2})}
= \fr{1}{2^{D-1}\pi^{D/2}\Gamma(\fr{D}{2})},
\en(19)
$$
where $R$ is a radius of the space or a size of system.
Thus the conformal kernels $\Box_D$ correspond to logarithmic Green functions
and  for this reason are connected with kernels of the field theories
equivalent to the generalized logarithmic gases \ci{16}.

An usual analysis (see for example \ci{17}) shows that a model (15)
does not have an usual phase transition with nonzero order parameter (OP)
due to logarithmic divergence of the OP fluctuations
$$
(\delta {\bf n}(x))^2 \sim \int d^D k G(k) \sim \int d^D k /k^D \sim \ln (R/a),
\en(20)
$$
where $\delta {\bf n}(x)$ is a deviation from some fixed value ${\bf n}_0.$
In this sense the D-dimensional conformal vector
NSM (15) are analogous
to 2D XY-model \ci{3,4}.
The thermodynamic properties of the model (15) will be
considered in other paper \ci{18}. Here we consider only its possible TE.
\bs

\cl{\bf III. Topological excitations with logarithmic energy.}

\bs

As was noted in \ci{12} an equation (16) can be represented in a "linear" form,
introducing new function $g(x):$
$$
\int \Box_D(x-x') {\bf n}(x') d^D x' = g(x) {\bf n}(x),
\en(21)
$$
$$
g(x) = \int ({\bf n}(x){\bf n}(x')) \Box_D(x-x') d^D x'.
\en(22)
$$
As all equations for NSM on spheres, it means that the action of the
operator $\Box_D$ on vector ${\bf n}(x)$ must be proportional to this vector, i.e.
the vector field ${\bf n}(x)$ must be in some sence an "eigenvector"
of the operator $\Box_D$ with
the "eigenvalue" $g(x)$, functionally depending on ${\bf n}(x).$
For each solution of equation (21) a value of the ${\cal S}[{\bf n}]$ can
be expressed through the "eigenvalue"
$$
{\cal S} = \fr{1}{2\alpha} \int g(x) d^D x
\en(23)
$$
Since $\pi_2(S^2)= \mbb {Z}$, there are the TE with topological charge
$Q \in \mbb {Z}$. The simplest TE with charge $Q=1,$ corresponding to the
identical map of spheres $S^2,$ must have the next asymptotic form
$$
n^i(x)_{r \gg a} \simeq  \fr{x^i}{r}
\en(24)
$$
Substituting (24) into equation (16), passing to the momentum space and using
the Fourier-image of the corresponding function
$$
n^i(k) = - i  \fr{C_D}{k^D} \; \fr{k^i}{k},
\en(25)
$$
where
$$
C_D = 2^{D-1} \pi^{\fr{D-1}{2}} (D-1) \Gamma((D-1)/2),
\en(26)
$$
we see, after returning to the $x$-space,
that the field (24) is an "eigenvector" of $\Box_D$ and, consequently,
a solution of equation (16).
The corresponding "eigenvalue" is
$$
g(x) = \fr{C^2_D}{(2\pi)^D r^D}.
\en(27)
$$
The action ${\cal S}$ of this solution can be found by two ways.
In the first case one can use (6,25) to obtain
$$
{\cal S} \simeq \fr{{\cal C}_D}{\alpha} \int \fr{dk}{k} f(ka)
\en(28)
$$
The integral in (28) is logarithmically divergent as it should be.
With a logarithmic accuracy one obtaines
$$
{\cal S} = \fr{{\cal C}_D}{\alpha} \ln (R/a), \quad
{\cal C}_D = \fr{C_D^2 S_{D-1}}{2(2\pi)^D}
= \fr{2^{D-2}\pi^{\fr{D-2}{2}}(D-1)^2 \Gamma^2(\fr{D-1}{2})}
{\Gamma(\fr{D}{2})}.
\en(29)
$$
The same result follows also from (23,27).
One can show that the interaction of two different TE with charges
$Q_1$ and $Q_2$ on large distancies
has a form of the Green function $G^D(r)$
$$
H_{12} (r) = Q_1 Q_2 G(r) \simeq  -  Q_1 Q_2 B_D  \ln (r/R)
\en(30)
$$
Analogously one can show that a field (24) is also
a solution of corresponding equations in even dimensional spaces.
For local kernel the equation is
$$
\Box^{2s} {\bf n}(x) -
{\bf n}(x) ({\bf n}(x)\Box^{2s}{\bf n}(x)) = 0, \quad D = 2s,
\en(31)
$$
and
$$
\Box^{2s} {\bf n}(x) = b_s/r^{2s} {\bf n}(x), \quad
b_s = \prod_{k=1}^s k(D-k) = \fr{\Gamma(D)\Gamma(\fr{D+2}{2})}
{\Gamma(D/2)}.
\en(32)
$$
Its energy is
$$
{\cal S} = \fr{b_s S_{D-1}}{2\alpha} \ln (R/a) = \fr{\pi^{D/2} \Gamma(D+1)}
{2\alpha \Gamma(D/2)}.
\en(33)
$$
For nonlocal kernel (11,12) the corresponding equation has the same form
as (16), but the "eigenvalue" $g(r)$ will now
contain a part with logarithmic factor, which gives for the energy
more complicated expression than simple logarithm. For this reason
the nonlocal kernels in even dimensional spaces are not so interesting ones.

Note that in the usual local D-dimensional NS-model with action
$$
{\cal S} = \fr{1}{2A} \int d^D x (\partial {\bf n})^2
\en(34)
$$
such TE have the energy
$$
E \simeq \fr{S_{D-1}(D-1)}{2 A(D-2)} (R^{D-2}-a^{D-2}).
\en(35)
$$
It is interesting that if we consider the mixed action ${\cal S}_{mix}$,
containing a sum of
kernels $\Box_d, \quad  d = 2,D,$
then the corresponding
equation has again the "hedgehog" solution (24) with total energy
$$
E = \fr{S_{D-1}(D-1)}{2 A(D-2)}(R^{D-2}-a^{D-2}) + \fr{{\cal C}_D}{\alpha} \ln (R/a).
\en(36)
$$
It means that a logarithmic part of the "hedgehog" energy can be
observed also in mixed models at scales
$$
(A/\alpha)^{1/(D-2)} > l > a.
$$
It is clear that analogous "anti-hedgehog" solutions with the same logarithmic
energies also exist.

\bs

\cl{\bf IV. Other topological excitations.}

\bs
Besides TE with logarithmic energy the TE of instantons type   with finite
energy can exist in the conformal NSM. They correspond to the configurations
with trivial boundary condition
$$
{\bf n}(x) \to {\bf n}_0, \quad r \to  \infty,
\en(37)
$$
where ${\bf n}_0$ is some constant unit vector. A necessary condition
for their existence is a nontriviality of other relevant homotopic group
$\pi_D(S^{D-1}).$ As is known from a general theory of homotopic groups of
spheres $\pi_D(S^{D-1})$ is defined through the suspension construction
by homotopic group $\pi_1(SO(D-1))$ \ci{15}. Since $\pi_1(SO(k)) = \mbb {Z}_2,
\; k>2,$ it follows from this that the conformal NSM on spheres
must have the instanton-like TE with topological charges
$$
Q \in \pi_1(SO(D-1)) = \mbb {Z}_2 = \mbb {Z} (mod 2) .
$$
All above topological charges are scalar (one-component).




The natural interesting generalization of the conformal NSM on spheres
is the D-dimensional conformal NS-models
on simple compact groups $G$ or on homogeneous spaces $G/H.$
As is known the nontrivial homotopic groups
of $G$ correspond to the so-called characteristic classes of this groups
\ci{15}
$$
\pi_k(G) \ne 0, \quad k = 2 k_i(G) - 1, \quad 1 \le i \le r(G)
\en(38)
$$
where $k_i(G)$ are the Weyl indices of group $G,$ and $r(G)$ is a rank
of group $G.$ It follows from (3) that
only odd homotopic groups can be nontrivial. All Weyl indices, the degree
of the Weyl invariant polynomials on the maximal abelian Cartan subalgebra
are known \ci{19}.
Then the TE with logarithmic energy are possible only for $D-1 = 2 k_i(G)-1,$
i.e. only for even dimensional spaces with $D = 2k_i(G)$, and the
instanton-like TE for odd $D = 2k_i(G) - 1.$
Since $\pi_k(G) = 0$ or  $\mbb {Z}$,
the corresponding topological charges will also be scalar ones.

In some cases it is very important to have a vectorial topological charges
\ci{20}.
A simple generalization of the sphere $S^{D-1}$, analogous to the torus
$T^n$ in 2D case, is a bouquet of $n$ spheres
$$
B_n^{D-1} = S^{D-1}_1\vee... \vee S^{D-1}_n
$$
and all spaces ${\cal M}$ with this first nontrivial topological cell
complex.
Then the topological charges will have  a vector form
$$
{\bf Q} = (q_1, ..., q_n), \quad q_i \in \mbb {Z}.
$$
But, in this case the  vector topological charges, corresponding to different
spheres,
will not interact between themselves due to their orthogonality
as in a case of torus \ci{20}.
For obtaining an interaction of topological charges in 3D case one
needs to consider NS-models on deformed $B^{D-1}_n$. For example,
the maximal flag spaces $F_G=G/T_G$
of the simple compact groups $G$ have $\pi_2(F_G)= \mathbb {L}_v,$
where $\mathbb {L}_v$ is a  dual root lattice of $G.$
Note that a sphere $S^2$ is a particular case of $F_G$:
$S^2 = SU(2)/U(1).$ Then the
3D conformal NS-models on $F_G$
will have topological excitations with a logarithmic energy
and interacting vector
topological charges ${\bf Q} \in \mathbb{L}_v.$
Since $\pi_3(F_G)= \pi_3(G)= \mbb {Z},$
in this case the "neutral" configurations will also have different
topological structure described by group $\pi_3(F_G).$

But for higher homotopic groups $\pi_i(F_G) = \pi_i(G),\quad i>3,$ thus
in higher dimensions ($D>3$) only scalar charges are possible
in these models.

\bs

This work was supported by RFBR grant No 96-15-96861.

\bs

\end{document}